# Mitigation of Saturated Cut-sets During Multiple Outages to Enhance Power System Security

Reetam Sen Biswas, *Student Member, IEEE*, Anamitra Pal, *Senior Member, IEEE*, Trevor Werho, *Member, IEEE,* and Vijay Vittal*, Life Fellow, IEEE*

*Abstract*—Ensuring reliable operation of large power systems subjected to multiple outages is a challenging task because of the combinatorial nature of the problem. Traditional approaches for security assessment are often limited by their scope and/or speed, resulting in missing of critical contingencies that could lead to cascading failures. This paper proposes a two-component methodology to enhance power system security. The first component combines an efficient algorithm to detect cut-set saturation (called the feasibility test (FT) algorithm) with real-time contingency analysis (RTCA) to create an integrated corrective action (iCA), whose goal is to secure the system against cut-set saturation as well as critical branch overloads. The second component only employs the results of the FT to create a relaxed corrective action (rCA) to secure the system against post-contingency cut-set saturation. The first component is more comprehensive, but the latter is computationally more efficient. The effectiveness of the two components is evaluated based upon the number of cascade triggering contingencies alleviated, and the computation time. The results obtained by analyzing different case-studies on the IEEE 118-bus and 2000-bus synthetic Texas systems indicate that the proposed two-component methodology successfully enhances the scope and speed of power system security assessment during multiple outages.

*Terms*— Power system security assessment, power system operations, multiple outages, saturated cut-set, graph theory, network flow algorithm

## I. Introduction

THE North American Electric Reliability Corporation (NERC) recommends that a reliable electric grid should be able to withstand the loss of a single element of its bulk power system (called *N*-1 reliability) [1]. Consequently, power system operators perform real-time contingency analysis (RTCA) and security constrained economic dispatch (SCED) successively at regular intervals [2]. RTCA evaluates the impact of a potential contingency on the system's static security. The critical contingencies detected by RTCA are modeled as security constraints in SCED to mitigate post-contingency overloads [3]. Despite RTCA-and-SCED trying to ensure *N*-1 reliability, cascading failures and blackouts/brownouts do occur in a power system. This highlights the need for additional/alternate means to enhance power system security.

Considering the high speed with which some of the blackouts propagate (the 2011 U.S. Southwest blackout occurred within 11 minutes [4]), a fast *and* robust assessment of power system security is imperative for real-time operations. For large systems, the traditional RTCA-SCED framework is not amenable to an *exhaustive N*-1 evaluation within a few minutes [5]. Therefore, power system operators have to select a *subset* of the contingencies for evaluation based upon some pre-defined criteria [6] or contingency ranking techniques [7]-[9]. This subset has considerable impact on RTCA performance: a large subset is computationally burdensome, while a small subset might miss critical scenarios [10].

After the 2011 U.S. Southwest blackout, the Federal Energy Regulatory Commission (FERC) reported that: "Affected TOPs (transmission operators) have limited visibility outside their systems, typically monitoring only one external bus. As a result, they lack adequate situational awareness of external contingencies that could impact their systems" [4]. However, modeling *all external events* will significantly increase the number of contingencies to be evaluated by the RTCA-SCED, thereby increasing the solution time considerably. Conversely, a *critical external event* that is not detected by RTCA will not be mitigated by SCED and could trigger cascading failures leading to unnecessary load-shedding.

Different network analysis schemes have been proposed recently to enhance situational awareness by performing *exhaustive N*-1 evaluation during multiple outages and screen out critical contingencies quickly [11]-[14]. In [11], Werho et al. used a network flow algorithm to identify the *cut-set* of minimum size between a source-sink pair; *a cut-set is a set of lines, which if tripped, would create disjoint islands in the network*. If the size of the minimum sized cut-set progressively decreased, it indicated a structural weakness between the selected source-sink pair. In [12], Beiranvand et al. used a topological sorting algorithm to screen out *coherent cut-sets*. A coherent cut-set is a set of lines that partitions the network, such that the power flows in the same direction through the lines. However, coherent cut-sets may not be the only bottlenecks in a power system. As such, in our prior research, we proposed a novel algorithm called the feasibility test (FT) algorithm to identify contingencies that create *saturated (or overloaded) cut-sets* [13]-[14]. Saturated cut-sets are the vulnerable interconnections of the system because they have limited power transfer capability. However, the research presented in [11]-[14] was limited to the *detection* of power system vulnerabilities. *How to quickly alleviate such vulnerabilities (saturated cut-sets) by corrective action(s) remained an open question!* When outages occur successively, it is important to (i) quickly identify as many critical problems as possible, and (ii) find solutions to the identified critical problems as quickly as possible. A two-component methodology is developed in this paper to attain these two objectives.

This work was supported in part by the National Science Foundation (NSF) Grant OAC 1934766 and the Power Systems Engineering Research Center (PSERC) Grant S-87.

The authors are associated with the School of Electrical, Computer, and Energy Engineering, Arizona State University (ASU), Tempe, AZ 85287, USA. (E-mail: rsenbisw@asu.edu, anamitra.pal@asu.edu, twerho@asu.edu, vijay.vittal@asu.edu).

Risk-based techniques for power system security assessment have also been developed [15]-[16]. These methods try to minimize the risk and uncertainty associated with possible contingencies that could manifest in the system. The proposed research is specifically aimed at minimizing the risk of *cascade triggering contingencies* that arise due to *post-contingency branch overloads* and *post-contingency cut-set saturation*, using a two-component methodology. The first component demonstrates how the detection and mitigation schemes for alleviating saturated cut-sets can be integrated with the traditional RTCA-SCED framework. As such, this component is focused on enhancing the scope of existing methods of power system security assessment. The second component proposes an alternative, computationally efficient approach to make power systems secure against saturated cut-sets quickly. The two components are implemented in parallel with the understanding that the solution of the second component be used only when the more comprehensive first component cannot provide a solution before the next redispatch occurs (see Section III.C for details on the real-time application).

## II. Theoretical Background

### A. Graph-theoretic terminologies in power system

The power system can be represented by a graph $\mathcal{G}(V, E)$, with buses contained in set $V$, and branches (transmission lines and transformers) contained in set $E$. The sets $G$ and $L$ contain all the generator (source) buses and load (sink) buses, respectively. Every transmission asset (line or transformer) is associated with a maximum power transfer capability referred to as the *asset rating*. Hence, every branch $e_l \in E$ is associated with a weight $f_l^{max}$, where $f_l^{max}$ denotes the asset rating of branch $e_l$. From the original graph $\mathcal{G}(V, E)$, two new graphs are created: the *flow graph*, $\mathcal{F}(V, E)$, and the *latent capacity graph*, $\mathcal{C}(V, E)$. The flow graph, $\mathcal{F}(V, E)$, contains information about the power flow through different branches of the network. Thus, if $f_l$ units of power flows through branch $e_l$ from bus $v_l^F$ towards bus $v_l^T$, a directed weight of $f_l$ is assigned to branch $e_l$ in a direction from $v_l^F$ to $v_l^T$. For the same branch $e_l$, the latent capacity graph, $\mathcal{C}(V, E)$, provides information regarding the extra flow that could be transferred from $v_l^F$ to $v_l^T$, and vice-versa. The bi-directional weights that denote the latent capacities of branches in $\mathcal{C}(V, E)$ are given by [14],

$$\left.\begin{array}{l} c_l^{FT} = f_l^{max} - f_l \\ c_l^{TF} = f_l^{max} + f_l \end{array}\right\} \quad (1)$$

where, $c_l^{FT}$ is the latent capacity in the direction from $v_l^F$ to $v_l^T$, and $c_l^{TF}$ is the latent capacity in the direction from $v_l^T$ to $v_l^F$. Moreover, since branch $e_l$ provides a direct connection from $v_l^F$ to $v_l^T$, it is called a *direct path* from bus $v_l^F$ towards bus $v_l^T$. Conversely, any set containing multiple branches joining bus $v_l^F$ to bus $v_l^T$ is called an *indirect path*.

### B. Detection of saturated cut-sets in power systems

A cut-set that transfers more power than is permitted by its maximum power transfer capability is called a *saturated* cut-set [13]. Mathematically, a cut-set $K$ would be a saturated cut-set if the following condition holds true:

$$F_K > R_K \quad (2)$$

where, $F_K = \sum_{\forall e_l \in K} f_l$ is the actual power flowing through cut-set $K$ and $R_K = \sum_{\forall e_l \in K} f_l^{max}$ is the maximum power that can be transferred across cut-set $K$ (limited by the asset ratings). The goal here is to screen out contingencies that will create a saturated cut-set in the system. If the outage of branch $e_l \in K$ overloads cut-set $K$, then the loss of $e_l$ would saturate cut-set $K$ by a margin of $R_K - F_K$ (called the transfer margin, $T_m$). Consider Fig. 1, in which cut-set $K^1$ contains three branches, i.e., $K^1 = \{e_4, e_6, e_7\}$. The total power transferred across this cut-set is $F_K^1 = f_4 + f_6 + f_7 = 360$ MW. The maximum power transfer capability of this cut-set is $R_{K^1} = f_4^{max} + f_6^{max} + f_7^{max} = 580$ MW. The cut-set $K^1$ is unsaturated as $F_{K^1} < R_{K^1}$. However, the loss of branch $e_4$ would saturate cut-set $K^1$. This is because with the outage of branch $e_4$, the power that must be transferred from Area 1 to Area 2 (assuming that the total load and generation remain the same) is still 360 MW (i.e., $F_{K^1} = 360$), but the maximum power transfer capability of cut-set $K^1$ reduces to 330 MW (as now $R_{K^1} = f_6^{max} + f_7^{max}$). Consequently, outage of branch $e_4$ will saturate cut-set $K^1$ by a margin of $R_{K^1} - F_{K^1} = 330-360 = -30$ MW.

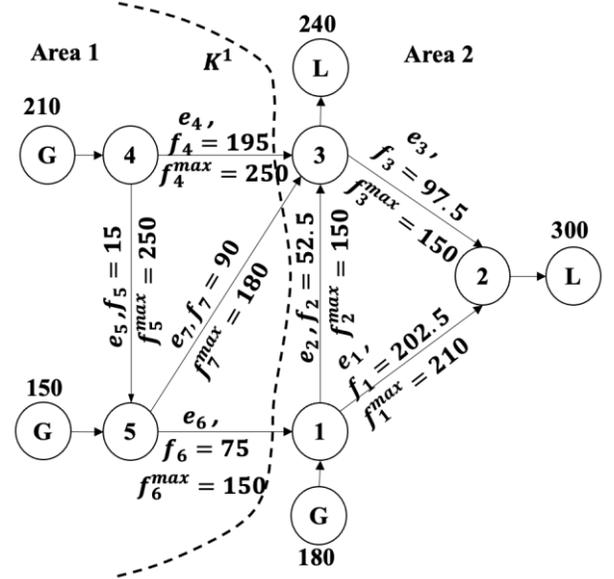

Fig. 1: Power transfer across a cut-set in the power system [13].

Now, a single branch can be associated with several cut-sets; e.g. branch $e_4$ of Fig. 1 is associated with cut-sets $K^2=\{e_4,e_5\}$, $K^3=\{e_1,e_2,e_4,e_7\}$, and $K^4=\{e_2,e_3,e_4,e_7\}$. Hence, evaluating the impact of a contingency on different cut-sets of a large power system is a computationally cumbersome task. A network flow algorithm (NFA) based on the law of conservation of energy was presented in [14] to find the contingencies that saturate a cut-set in the network at an enhanced computation speed. The important aspects of the NFA (in the context of this paper) are summarized below:

1. The NFA was based on the following principle: utilize available generation of sources (generators) to satisfy total demand of sinks (loads), without violating any asset ratings. Although the flow solution was non-unique, it was proved in [14] that for detecting saturated cut-sets any valid flow solution was sufficient.
2. A *feasibility test* (FT) algorithm was developed to uniquely determine if a contingency saturated a cut-set in the network. The FT employed exhaustive breadth first search (BFS) [17] graph traversal scheme to find the total power transfer capability of the set of indirect paths of a branch. If the set of indirect paths did not have sufficient capacity to re-route the power flowing through the branch (direct path), it meant that the outage of that branch saturates a cut-set in the network; the branch was termed a *special asset*. If the outage of a branch saturated more than one cut-set, the FT screened out the cut-set that was saturated by the largest margin; this cut-set was called the *limiting critical cut-set, $K_{crit}$*. The FT also provided the *transfer margin, $T_m$*, by which $K_{crit}$ got saturated.

3. A computationally efficient *update scheme* (UPS) was proposed to update $\mathcal{F}(V, E)$ quickly when a branch outage occurred. By efficiently rerouting flows along the indirect paths of the branch that suffered the outage, the UPS circumvented the need for re-building from scratch the weighted flow graph of the network following the outage.
4. A *shortlisting assets* (SA) algorithm was developed to screen out the subset of contingencies that must be re-evaluated by FT following an outage. This algorithm exploited the information provided by FT performed on the pre-outage system and used the updated $\mathcal{F}(V, E)$ obtained by UPS to speed up the computations.

More information about this NFA can be found in [14].

### C. Practical aspects of the FT algorithm

The FT algorithm *guarantees* detection of all contingencies that create saturated cut-sets but not all post-contingency branch overloads are identified [14]. Therefore, modeling the cut-set power transfer constraints introduced by the FT in the dispatch will ensure that the resulting solution is at least secure against all *post-contingency cut-set saturation*. As saturated cut-sets are the "vulnerable bottlenecks within power grids and represent seams or fault lines across which islanding seems likely" [12], securing the system against all post-contingency cut-set saturation is an important achievement. In this paper, the FT has been used to detect saturated cut-sets based on thermal ratings of the assets. However, as mentioned in [14], the FT can also incorporate ratings from other analyses (such as proxy limits based on power system stability criteria).

### D. RTCA-SCED for real-time power system operations

RTCA and SCED are usually employed by power system operators to operate the system in a secure manner. Fig. 2 shows a schematic of the traditional RTCA-SCED framework that takes its inputs from the state estimator. SCED finds a least cost redispatch solution to eliminate the potential post-contingency branch overloads identified by RTCA. The solution obtained by SCED is fed back into RTCA to ensure that the new solution does not create additional overloads. When no additional violations are detected, the redispatch solution is implemented in the power system.

In practice, only a subset of all possible branch overloads is fed as inputs to RTCA; these selected critical branches form the *contingency list* [10]. This contingency list is determined from offline studies [3], operator knowledge [18], or contingency ranking techniques [7]-[9]. As the contingency list is *not exhaustive*, it is possible that an important contingency is left out from this list, due to which it is not detected by RTCA (and hence *not* corrected by SCED) until it is too late. This is a serious limitation especially during extreme event scenarios when successive outages occur quickly. Furthermore, when multiple outages have already occurred, a larger number of post-contingency overloads manifest, because the system is in a stressed operating condition. Therefore, SCED takes a longer time to find a solution due to the increased number of security constraints that it has to model. Different rounding conventions of distribution factors and approximations have also been applied to the SCED model to enhance its computational speed [19]. However, the increased solution time under extreme exigencies might encourage use of larger approximations, which would then affect the solution quality. Thus, both the *scope* as well as the *speed* of traditional power system security assessment must be enhanced during multiple outage scenarios.

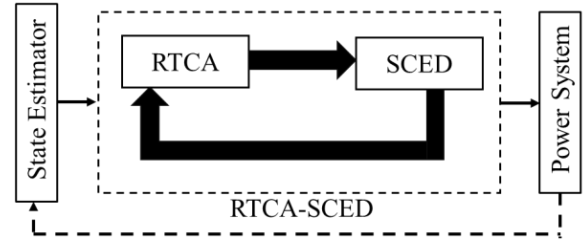

Fig. 2: RTCA-SCED for real-time power system operations

## III. Two Component Methodology

The proposed methodology consists of two components. The first component increases the scope of the traditional RTCA-SCED framework so that it not only focuses on selected branch overloads, but also saturated cut-sets. The second component speeds up the response time by only alleviating the saturated cut-sets, which by their very definition capture the more congested regions of the system.

### A. The first component of proposed methodology

The proposed *first component* aims to make the power system secure against post-contingency cut-set saturation as well as critical branch overloads by integrating the results from FT and RTCA to create an integrated corrective action (iCA), as shown in Fig. 3. The objective of the iCA is to find a least cost redispatch solution to ensure that the critical contingencies detected by RTCA do not create post-contingency branch overloads, and the special assets identified by FT do not create saturated cut-sets. During multiple outage scenarios, it is possible that a generation redispatch alone is not able to mitigate all the identified overloads. Under such circumstances, controlled load-shedding will be implemented. Since disconnecting the loads incur high socio-economic costs [20], load-shedding will be used as the last resort during redispatch.

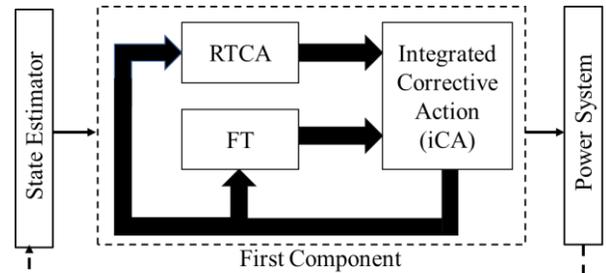

Fig. 3: The first component: The results from RTCA and FT are used to create an integrated corrective action (iCA)

Consider that the generator at bus $i \in G$ in the system is associated with a quadratic cost curve as shown below:
$$F_i(G_i) = a_i + b_i G_i + c_i G_i^2 \quad (3)$$
where, $G_i$ is the power produced (in MW) by the generator at bus $i$, and $a_i$, $b_i$, and $c_i$ are the fixed cost coefficient (in \$), the linear cost coefficient (in \$/MW), and the quadratic cost coefficient (in \$/MW$^2$), respectively, for the corresponding generator. Let $G_i^o$ and $G_i^n$ denote the power produced before and after the new dispatch. The change in generation cost as a function of change in power generation, $\Delta G_i (= G_i^n - G_i^o)$, is given by,
$$\Delta F_i(\Delta G_i) = \{a_i + b_i G_i^n + c_i (G_i^n)^2\} - \{a_i + b_i G_i^o + c_i (G_i^o)^2\}$$
$$= c_i \Delta G_i^2 + d_i \Delta G_i \quad (4)$$
where, $d_i = (2c_i G_i^o + b_i)$. Now, the cost of shedding the load at bus $j \in L$ can be written as:
$$\Delta F_j(\Delta L_j) = m_j \Delta L_j \quad (5)$$
where, $\Delta L_j$ denotes the amount of load-shed, and $m_j$ is the cost coefficient of load-shed (in \$/MW); $m_j$ is chosen to be significantly higher compared to the generator cost coefficients because the goal is to use load-shed only when generation redispatch alone cannot mitigate all violations. The optimization problem that minimizes the total cost of change in generation and load-shed can now be written as:
$$\text{Minimize:} \sum_{i \in G} (c_i \Delta G_i^2 + d_i \Delta G_i) + \sum_{j \in L} (m_j \Delta L_j) \quad (6)$$
The constraints applied to (6) are as follows.

*1. Branch power flows*

To model the branch power flow limits, the power transfer distribution factors (PTDFs) are used. PTDFs are linear sensitivity factors that approximate the change in flow through a line caused by a change in power injection in the system [21]. Let, $PTDF_{l,i}$ denotes the change in flow in branch $e_l$, for one unit of power added at bus $i$ and one unit of power withdrawn from the reference bus of the system. Then, the change in flow, $\Delta f_l$, through $e_l$ for the change in bus power injections can be obtained as follows:
$$\Delta f_l = \sum_{\forall i \in G} PTDF_{l,i} \Delta G_i - \sum_{\forall j \in L} PTDF_{l,j} \Delta L_j \quad (7)$$
Consequently, the constraint equation for the maximum and minimum power flows is given by:
$$f_l^{min} - f_l^0 \le \sum_{\forall i \in G} PTDF_{l,i} \Delta G_i - \sum_{\forall j \in L} PTDF_{l,j} \Delta L_j$$
$$\le f_l^{max} - f_l^0, \quad \forall e_l \in E \quad (8)$$
where, $f_l^o$, $f_l^{max}$ and $f_l^{min}$ denote the original power flow, maximum and minimum power flow limits, respectively.

*2. Power injections*

The maximum and minimum power generation constraint is given as follows:
$$G_i^{min} - G_i^0 \le \Delta G_i \le G_i^{max} - G_i^0, \quad \forall i \in G \quad (9)$$
where, $G_i^o$, $G_i^{max}$, and $G_i^{min}$ denote the original power produced, maximum power and minimum power that can be produced by the generator at bus $i$, respectively. Similarly, the constraints for minimum and maximum power demand at a load bus $j$ is given as follows:
$$L_j^{min} - L_j^0 \le \Delta L_j \le L_j^{max} - L_j^0, \quad \forall j \in L \quad (10)$$

*3. Security constraints 1: Post-contingency branch flows*

The post-contingency branch flow constraints can be efficiently modeled using the line outage distribution factors (LODFs) [21]. Let $LODF_{l,k}$ represent the change in flow through branch $e_k$ that will appear on branch $e_l$ for an outage of branch $e_k$. The post-contingency flow, $f_l^c$, through $e_l$ for a potential outage of $e_k$ is given as follows:
$$f_l^c = f_l^n + LODF_{l,k} f_k^n \quad (11)$$
where, $f_l^n$ and $f_k^n$ denote the new flows corresponding to the new redispatch solution through branches $e_l$ and $e_k$, respectively. Now, (11) can be re-written as:
$$f_l^c = (f_l^0 + \Delta f_l) + LODF_{l,k}(f_k^0 + \Delta f_k) \quad (12)$$
where, $f_l^0$ and $f_k^0$ denote the original flows through branches $e_l$ and $e_k$, respectively, and $\Delta f_l$ and $\Delta f_k$ represent the incremental change in flows through branches $e_l$ and $e_k$ as obtained from the redispatch. Substituting $\Delta f_l$ and $\Delta f_k$ from (7) into (12), and using the respective branch flow limits, we obtain the following equations for post-contingency branch flow constraints:
$$f_l^{min} - (f_l^0 + LODF_{l,k} f_k^0)$$
$$\le \sum_{\forall i \in G} (PTDF_{l,i} + LODF_{l,k} PTDF_{k,i}) \Delta G_i -$$
$$\sum_{\forall j \in L} (PTDF_{l,j} + LODF_{l,k} PTDF_{k,j}) \Delta L_j \le f_l^{max} -$$
$$-(f_l^0 + LODF_{l,k} f_k^0), \quad \forall e_k \in E_v, \forall e_l \in E \quad (13)$$
where, set $E_v$ contains the critical contingencies detected by RTCA. Equation (13) is modeled for all post-contingency overloads for the critical contingencies detected by RTCA.

*4. Security constraints 2: Cut-set power transfer*

This type of security constraint is designed for the special assets detected by the FT algorithm [14]. The objective here is to reduce the total power transfer across the limiting critical cut-set $K_{crit}$ by the respective transfer margin, $T_m$, as follows:
$$\sum_{\forall e_l \in K_{crit}} \Delta f_l \le T_m, \quad (14)$$
Now substituting $\Delta f_l$ from (7) into (14), the constraint for cut-set power transfer is obtained as follows:
$$\sum_{\forall i \in G} \left( \sum_{\forall e_l \in K_{crit}} PTDF_{l,i} \right) \Delta G_i -$$
$$\sum_{\forall j \in L} \left( \sum_{\forall e_l \in K_{crit}} PTDF_{l,i} \right) \Delta L_j \le T_m, \forall K_{crit} \in \mathcal{K}_{crit} \quad (15)$$
where, the set $\mathcal{K}_{crit}$ contains the limiting critical cut-sets detected by the FT corresponding to different special assets.

Note that a SCED can essentially solve the same optimization problem as iCA with all the constraints modeled except the cut-set power transfer constraints [22]. By considering both post-contingency branch overloads as well as post-contingency cut-set saturation, the iCA creates a more comprehensive corrective action than SCED.

*B. The second component of proposed methodology*

The first component of Section III.A (or the traditional RTCA-SCED framework of Section II.C) is likely to take more time because of the larger number of security constraints modeled in the optimization problem for iCA (or SCED). For example, if the number of critical contingencies detected by RTCA is $|E_v|$, and the total number of transmission assets is $|E|$, the number of post-contingency branch flow constraints (see security constraints 1) to be modeled is $|E_v| \times |E|$. For a large power system, containing thousands of branches, $|E|$ is large. Moreover, for a stressed power system that has suffered multiple outages, $|E_v|$ is also large. Consequently, the proposed first component (or RTCA-SCED) may not be able to suggest corrective actions at high speeds.

To provide a high-speed corrective action, a second component is proposed, which only utilizes the results from FT to create a relaxed corrective action (rCA) as shown in Fig. 4. The rCA solves the same optimization problem (given by (6)), but *without modeling the post-contingency branch flow constraints* (described by (13)). However, the cut-set power transfer constraints, described by (15), are retained in rCA, i.e., the rCA utilizes the results from FT to only secure the system against post-contingency cut-set saturation. Note that if the optimization problem given by (6) is solved without modeling any security constraints (neither (13) nor (15)), it can reduce to an optimal power flow (OPF) problem [23]. Therefore, by considering (15), the rCA adds a relaxed criterion of power system security onto an OPF problem.

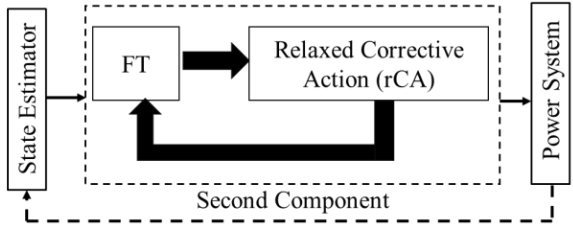

Fig. 4: The second component: The results from FT are only utilized to create a relaxed corrective action (rCA)

If the set $E_s$ contains the special assets detected by FT, the number of cut-set power transfer equations modeled by the rCA is $|E_s|$. Now, as the number of cut-set violations identified will be smaller than the total number of branches of a power system, $|E_s| \ll |E|$, and consequently, $|E_s| \ll |E_v| \times |E|$. *This implies that the number of security constraints modeled by the rCA is significantly less compared to the number of security constraints modeled by the iCA (or SCED)* and is the primary reason for the very high speed of rCA.

It should however be noted that the solution obtained using the second component is secure against pre-contingency branch overloads and post-contingency cut-set saturation, but not post-contingency branch overloads. Conversely, the solution obtained from the first component is secure against post-contingency cut-set saturation, as well as pre-contingency and post-contingency branch overloads. Naturally, the solution quality of the first component is better than the second.

At the same time, it is important to note that if generation redispatch alone cannot provide a feasible solution with respect to a relaxed set of constraints such as those used in rCA, it is obvious that generation redispatch will not provide a solution with more comprehensive constraints such as those used in iCA. Therefore, if load-shedding is indicated by rCA (in the second component), it will also be indicated by iCA (in the first component); albeit after a longer time and the amount of load-shed will be equal or higher. Therefore, the ability to quickly indicate the minimum amount of load that must be shed before a detailed network analysis tool can provide a more accurate estimate of load-shed, is another advantage of the rCA.

*C. Real-time application of the proposed two-component methodology*

It can be discerned from Sections III.A and III.B that the first and second components enhance the scope and speed, respectively, of traditional power system security assessment. The question then becomes, *how should the two components be applied in real-time when a contingency occurs?* Different entities implement SCED at different timescales for real-time power system operations. For example, PJM Interconnection LLC implements real-time SCED every fifteen minutes [24], whereas Midcontinent Independent System Operator (MISO) implements SCED every five minutes [25]. In this context, the real-time application of the two components can be explained with the help of the timelines shown in Fig. 5.

With reference to Fig. 5, let an outage occur at time $t_o$. After the outage, the first and second components should be initiated simultaneously but independently. Let the redispatch solution be implemented at time $t_d$, while the first and second components provide their dispatch solutions at time $t_i$ and $t_r$, respectively. If $t_i < t_d$, as shown in Fig. 5(a), then the solution obtained using the first component should be used for redispatch as it has better quality. However, if $t_i > t_d$ and $t_r < t_d$ as shown in Fig. 5(b), then the solution obtained from the second component should be implemented to at least secure the system against post-contingency cut-set saturation. It will be shown in Section IV.B.1 that the computational speed of the second component is comparable to a DC-OPF. As such, the likelihood of $t_r > t_d$ is small even for large power systems. However, if that still happens then depending on its availability, the solution from the first (preferred) or the second component should be implemented in the next redispatch.

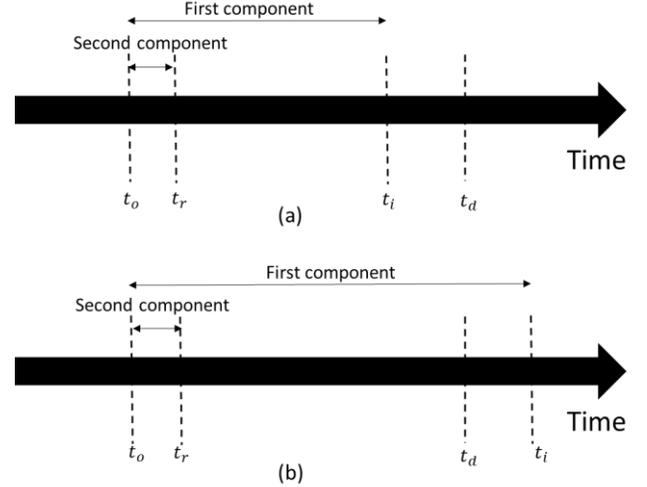

Fig. 5: (a) If the first component provides a dispatch solution before the scheduled time for the next redispatch, then the solution obtained from the first component should be implemented. (b) If the first component does not provide a dispatch solution before the scheduled time for next redispatch, then the solution obtained from the second component should be implemented.

*D. Modified Update Scheme (M-UPS)*

The corrective actions introduced by iCA (in the first component) and rCA (in the second component) change the bus power injections. Therefore, FT must re-evaluate the system corresponding to the new bus power injections to ensure that the updated system does not have any additional saturated cut-sets due to a potential outage. Hence, a modified-update scheme (M-UPS) is developed in this paper that updates the flow and latent capacity graphs in a computationally efficient manner, thereby eliminating the need for recreating these weighted graphs from scratch. Let the sets $V^p$ and $V^n$ contain the buses where power injection has increased and decreased, respectively. Similarly, let $\Delta I_p$ and $\Delta I_n$ denote the increase and decrease in net power injection at buses $v_p \in V^p$ and $v_n \in V^n$, respectively. Now the updated flow and latent capacity graphs can be obtained using **Algorithm I**.

**Algorithm I**: Modified Update Scheme (M-UPS)
i. Randomly select a source $v_p \in V^p$ and a sink $v_n \in V^n$.
ii. Search $\mathcal{C}(V, E)$ to traverse the shortest unsaturated path $P$ from $v_p$ to $v_n$ using breadth first search (BFS) [17].
iii. Use $\mathcal{C}$ to find the maximum extra flow, $C_P$, that can be transferred from $v_p$ to $v_n$ through path $P$.
iv. Obtain the flow, $F_P$, to be injected in $\mathcal{F}(V, E)$ along path $P$ from $v_p$ to $v_n$ as $F_P = \min(\Delta I^p, \Delta I^n, C_P)$.
v. Update weights of branches in graph $\mathcal{F}$ as $f_l = f_l + F_P$, and in graph $\mathcal{C}$ as per (1), for all branches that belong to path $P$.
vi. Update net power injections at $v_p$ and $v_n$ as $\Delta I^p \coloneqq \Delta I^p - F_P$ and $\Delta I^n \coloneqq \Delta I^n - F_P$.
vii. Depending upon the values of $\Delta I^p$ and $\Delta I^n$, update the source and sink in accordance with the following logic:
 a. if $\Delta I^p \neq 0$ & $\Delta I^n \neq 0$, the source and sink are not changed.
 b. if $\Delta I^p = 0$ & $\Delta I^n \neq 0$, a new source $v_p$ is selected from set $V^p$, keeping the sink $v_n$, unchanged.
 c. if $\Delta I^p \neq 0$ & $\Delta I^n = 0$, a new sink is selected from set $V_n$, keeping the source $v_p$, unchanged.
viii. Repeat Steps (ii) through (vii) until the total increase in power injection is compensated by the total decrease in power injection.

The M-UPS uses the set of shortest indirect paths that have extra capacity to re-route the flows. The reason for this is explained with the help of an example. Consider the 5-bus system of Fig. 1. The flow solution in Fig. 1 is a DC power flow solution. A graph-theory based network flow solution of the same system obtained using the NFA [14] is shown in Fig. 6. Detailed information on the creation of this flow solution can be found in [13]. Fig. 7(a) and Fig. 7(b) present the power transfer across cut-set $K_1$={4-3,5-3,5-1} for the flow graphs of Fig. 1 and Fig. 6, respectively. Despite the individual branch flows being different, the total power transfer across cut-set $K_1$ is 360 MW. If the FT is applied on any of the flow graphs, it will detect that the outage of branch 4-3 saturates cut-set $K_1$ by 30 MW. This is because the total power transfer capacity of cut-set $K_1$ will reduce to 330 MW upon outage of branch 4-3. *Hence, for detecting saturated cut-sets, the net power transfer across any cut-set of the network is important, rather than the individual branch flows.* Since it does not matter which paths are selected to match the total load with generation, following a system redispatch, the set of shortest indirect paths can be used to re-route the flows using **Algorithm I**. A rigorous analysis of this subject can be found in [13] and [14].

Moreover, utilizing the set of shortest indirect paths among the buses where the bus power injections have changed implies that a small sub-graph of the network is used to create the updated graphs. This not only enhances the computational efficiency of creating an updated flow graph, but also helps in shortlisting the assets (explained in more detail in Section III.E) to be re-evaluated by the FT following the redispatch.

*E. Modified Shortlisting Assets (M-SA)*

In the pre-outage scenario, all assets are evaluated by the FT. However, once the M-UPS creates an updated flow graph it may not be necessary to test all assets by the FT once again to identify the set of special assets. Hence, a modified-shortlisting asset (M-SA) scheme is developed in this paper which finds the contingencies to be evaluated by FT following the update of the flow graph to account for the changes in bus power injections.

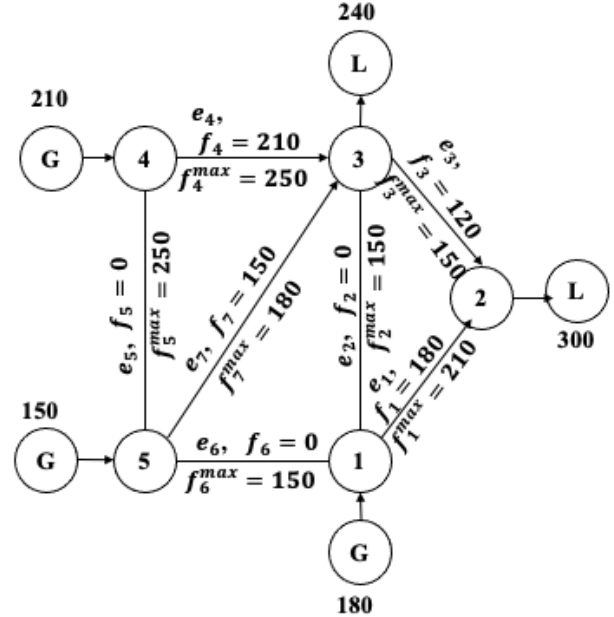

Fig. 6: Another graph-theory based network flow solution of the same 5-bus system of Fig. 1 [13].

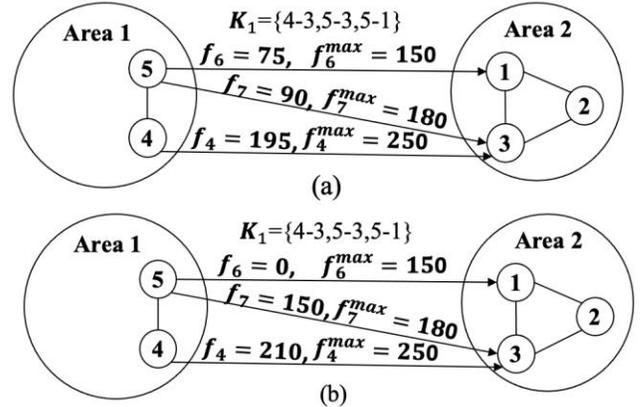

Fig. 7: (a) Power transfer across cut-set $K_1$ for the flow graph of Fig. 1, (b) Power transfer across cut-set $K_1$ for the flow graph of Fig. 6.

The concept of M-SA is explained with the help of Fig. 8. Let the M-UPS modify the flows through path $P_2$ in the network to account for the changes in bus power injections. Also, from the FT performed in the pre-outage scenario, let it be known that the flow of another branch $e_m$ can be re-routed through path $P_1$. Now, if paths $P_1$ and $P_2$ do not have any common branches as shown in Fig. 8(a); FT need not be repeated for branch $e_m$. This is because it is already known from the pre-outage scenario analysis that the outage of $e_m$ does not saturate a cut-set and the disrupted flow can be re-routed through path $P_1$ itself. However, if paths $P_1$ and $P_2$ have branches in common as shown in Fig. 8(b), then $e_m$ must be re-evaluated by FT, once the network flows have been updated.

Table I: Comparative analysis of the first component and RTCA-SCED for a sequence of six outages in the IEEE 118-bus test system

| Event (branch outages) | First component (FT-RTCA-iCA) | | | | | RTCA-SCED | | | |
|---|---|---|---|---|---|---|---|---|---|
| | FT | RTCA | MATCASC (before correction) | Gen. Cost (k$) | MATCASC (after correction) | RTCA | MATCASC (before correction) | Gen. Cost (k$) | MATCASC (after correction) |
| Outage 1: 15-33 | - | - | - | 126.2 | - | - | - | 126.2 | - |
| Outage 2: 19-34 | - | 5-8 | - | 126.3 | - | 5-8 | | 126.3 | - |
| Outage 3: 37-38 | 42-49 | 42-49, 5-8, 26-30 | 42-49 | 126.5 | | 42-49 5-8, 26-30 | 42-49 | 126.5 | |
| Outage 4: 42-49 | 45-46, 45-49 | 45-46, 45-49 | 45-46, 45-49 | 126.7 | | 45-46, 45-49 | 45-46, 45-49 | 126.7 | - |
| Outage 5: 49-66 | - | 5-8 | - | 126.7 | - | 5-8 | - | 126.7 | - |
| Outage 6: 66-67 | 64-65, 65-66 | 64-65 | 64-65, 65-66 | 127.1 | - | 64-65 | 64-65, 65-66 | 126.9 | 65-66 |

Table II: Comparative analysis of the second component and DC-OPF for a sequence of six outages in the IEEE 118-bus test system

| Event (branch outages) | Second component (FT-rCA) | | | | DC-OPF | |
|---|---|---|---|---|---|---|
| | FT | MATCASC (before correction) | Gen. Cost (k$) | MATCASC (after correction) | Gen. Cost (k$) | MATCASC |
| Outage 1: 15-33 | - | - | 126.2 | - | 125.9 | 26-30 |
| Outage 2: 19-34 | - | - | 126.2 | - | 125.9 | 26-30 |
| Outage 3: 37-38 | 42-49 | 42-49 | 126.3 | - | 125.9 | 26-30, 42-49 |
| Outage 4: 42-49 | 45-46, 45-49 | 45-46, 45-49 | 126.4 | - | 126.2 | 26-30, 45-46 42-49 |
| Outage 5: 49-66 | - | - | 126.4 | - | 126.2 | 26-30, 45-46 45-49 |
| Outage 6: 66-67 | 64-65, 65-66 | 64-65, 65-66 | 126.7 | 64-65 | 126.2 | 26-30, 45-46, 45-49, 64-65, 65-66 |

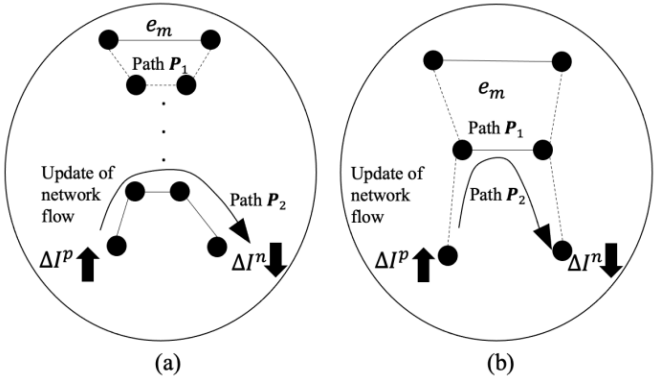

Fig. 8: (a) Updating the flows in the network for a change in the power injections does not involve any branch in the indirect paths of branch $e_m$; (b) Updating the flows in the network for a change in the power injections involves branches in the indirect paths of branch $e_m$.

Note that the M-UPS and M-SA are used to perform FT successively when the corrective actions change the bus power injections. Conversely, the UPS and SA developed in [14] were used to perform FT successively following a branch outage that had occurred in the system.

## IV. SIMULATION RESULTS

### A. IEEE 118-bus test system

We initially present the performance of the proposed two-component methodology against traditional approaches, such as RTCA-SCED or DC-OPF, using a detailed case-study that involves a sequence of six outages. Subsequently, to demonstrate consistency, its performance is compared with the traditional approaches for 40 additional case-studies. All simulations were done in MATLAB. GUROBI was used to solve the optimization problems.

*1) A detailed case-study of the IEEE 118-bus test system*

The performance of the first component is presented and compared with the RTCA-SCED framework when six outages manifest successively. The first column of Table I shows the sequence of events. Columns two through six present the results associated with the first component. The second column shows the special assets detected by the FT. An outage of any of these special assets (after the outage that has already occurred in the corresponding row of the first column), will create post-contingency cut-set saturation. The third column shows the critical contingencies detected by RTCA that result in post-contingency branch overloads. To determine the entries of this column, a two-step procedure was followed: (a) PTDFs and line ratings were used to rank the contingencies following every outage [9], and (b) top 30% of the contingencies [5] were evaluated by RTCA to determine the post-contingency branch overloads. The special assets detected by FT in the second column and the critical branch contingencies detected by RTCA in the third column were set as inputs to the iCA.

Next, an independent cascading simulation analysis was conducted using MATCASC [26], a software package that evaluates the consequence of cascading failures in power systems. To screen out outages that will trigger a cascade and result in unserved power demand, every outage was evaluated by MATCASC. The fourth and sixth columns of Table I present the cascade triggering contingencies detected by MATCASC *before* and *after* the implementation of iCA. The fifth column presents the redispatch solution (generation cost) obtained from the iCA. Note that the redispatch solution for this case-study did not result in any load-shed. Finally, we observe from the sixth column that the solution obtained from iCA does not contain any cascade triggering contingencies. *Therefore, through iCA, the first component has effectively utilized the* *information from FT and RTCA to mitigate the risk of cascade triggering contingencies for the given sequence of events.*

Now, we evaluate the performance of the RTCA-SCED framework for the same sequence of events. Columns seven through ten of Table I present the results associated with RTCA-SCED. The column headings are similar to that of the first component, with the exception that the FT results are absent in this section as the traditional SCED only utilizes the inputs from RTCA. For the first five outages the results of the first component and RTCA-SCED are identical. This is because for these outages the FT does not identify additional violations to those already detected by RTCA (compare the second and third columns of Table I). However, after the sixth outage FT detects the special asset 65-66 in addition to the critical contingency 64-65 identified by RTCA (see second and third column of the last row). This becomes the basis for the difference in the redispatch solutions of the first component and RTCA-SCED as seen in the fifth and ninth columns of the last row. Finally, it is observed that the RTCA-SCED solution contains one cascade triggering contingency (65-66), while the solution obtained from iCA did not have any (compare sixth and tenth columns of the last row). *This observation proves that integrating the results from FT with RTCA enhances the ability of power system security assessment in mitigating the risk of cascade triggering contingencies.*

Now, there could be situations when the first component takes a long time to generate a solution, in which case the second component should be utilized as discussed in Section III.C. Table II presents the application of the second component and compares it with a simple DC-OPF. Note that it is fair to compare the second component with a DC-OPF instead of an AC-OPF because the DC-OPF solves a linearized constrained optimization problem (similar to rCA used in the second component) while the optimization problem solved in AC-OPF is non-linear. Moreover, the focus here is on high-speed, and it is well-known that for any given system, a DC-OPF problem can be solved much faster than an AC-OPF problem.

The first column of Table II lists the sequence of events. Columns two through five present the results of the second component. Note that only the FT results are shown in this section as the RTCA results are not considered in the second component. Cascading analysis done after the corrective action indicates that the redispatch obtained from rCA does not contain any cascade triggering contingency for the first five consecutive outages (see fifth column of Table II). However, after the sixth outage, two cascade triggering contingencies manifest before the corrective action is initiated (see last row, third column of Table II), of which, only one is addressed by rCA. That is, the solution obtained using the rCA still contains one cascade triggering contingency (see last row, fifth column of Table II). This happened because the contingency 64-65 triggered cascading failures due to branch overloads, even after the rCA alleviated all post-contingency cut-set saturation.

However, the second component performs significantly better than a DC-OPF (see columns six and seven of Table II). The sixth column presents the DC-OPF redispatch solution, while the seventh column presents the results of the cascading analysis by MATCASC on the redispatch solution. Since a DC-OPF does not model any security constraints, the number of cascade triggering contingencies in its solution is significantly high in comparison to the one obtained using rCA (in the second component). *This shows that in situations when the first component takes a long time to generate a solution due to heavy computational burden, the second component should be used to secure the system against post-contingency cut-set saturation, and thereby reduce the risk of cascading failures.*

*2) Comparison of the proposed methodology for different case-studies in the IEEE 118-bus test system*

To validate the consistency of the two-component methodology, 40 different case-studies were generated (in addition to the case-study presented in detail in Section IV.A.1). To produce critical scenarios, multiple successive outages were created in different regions of the system. The number of successive outages varied from two to six for the different case-studies. The redispatch solution obtained from the proposed and traditional approaches were evaluated by MATCASC to check if the solution contained cascading contingencies for any of the outages involved in the case-study.

As the computation time of the first component and the traditional RTCA-SCED framework are of similar order (verified experimentally in Section IV.B.1), their performance, denoted by bars with A and B markers, respectively, in Fig. 9, were compared first. It is observed from the figure that the redispatch solution from RTCA-SCED contained cascade triggering contingencies for case-studies involved with three (1), four (2), and six (1) outages. However, when the first component was used, none of the case-studies contained any cascade triggering contingencies (bar A is absent in Fig. 9).

Owing to the similar computation time of the second component and DC-OPF (verified experimentally in Section IV.B.1), their performance, denoted by bars C and D, respectively, in Fig. 9, were compared next. It is observed from the figure that the redispatch solution from DC-OPF contained cascade triggering contingencies for all 41 case-studies. This is because a DC-OPF does not model any security constraints. However, when the second component was used, the number of case-studies containing cascade triggering contingencies decreased considerably in comparison to the DC-OPF results (compare the heights of bars C and D in Fig. 9). *This statistical comparison confirms that during multiple outage scenarios, the proposed two-component methodology can lower, if not eliminate, the risk of cascade triggering contingencies in comparison to traditional approaches.*

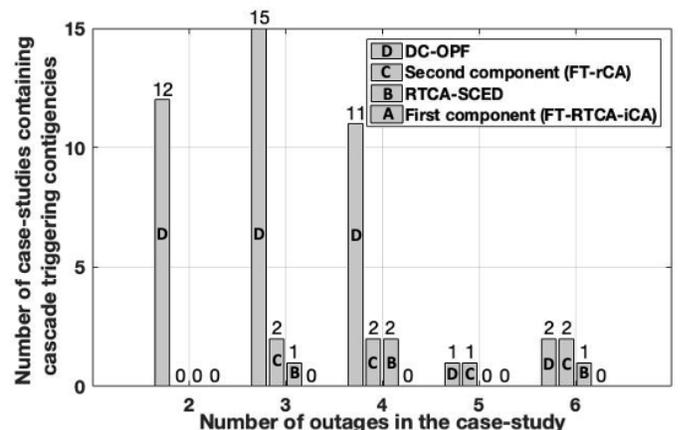

Fig. 9: Statistical summary of the performance of different approaches for 41 case-studies in the IEEE 118-bus test system

**Table III: Time comparisons of different approaches for the 2000-bus synthetic Texas system**

| Event | Time* | | | |
|---|---|---|---|---|
| | RTCA-SCED | First component (FT-RTCA-iCA) | DC-OPF | Second component (FT-rCA) |
| Outage 1: 3047-3129 | 388 sec | 421 sec | 15 sec | 28 sec |
| Outage 2: 1004-3133 | 431 sec | 487 sec | 20 sec | 21 sec |
| Outage 3: 3127-3141 | 622 sec | 720 sec | 24 sec | 20 sec |

* The simulations were performed on a computer with 2.3 GHz Dual-Core Intel Core i5 processor and 8 GB RAM.

**Table IV: Real-time application of the two-component methodology on the 2000-bus synthetic Texas System**

| Event | Method | Dispatch Solution | | No. of cascade triggering contingencies detected by MATCASC (after correction) |
|---|---|---|---|---|
| | | Gen. cost (k$) | Load-shed (MW) | |
| Outage 1: 3047-3129 | First component: FT-RTCA-iCA | 922.4 | 106 | 0 |
| Outage 2: 1004-3133 | First component: FT-RTCA-iCA | 924.7 | 0 | 0 |
| Outage 3: 3127-3141 | Second component: FT-rCA | 923.2 | 0 | 0 |

### B. Texas 2000-bus test system

The scalability, computation time, and real-time implementation of the proposed methodology are demonstrated for a case-study of the 2000-bus synthetic Texas system [27]. The total power demand in the system is 67,109 MW.

#### 1) Computation times of different approaches

Consider a sequence of three successive outages in this system occurring at intervals of 10 minutes each, as shown in the first column of Table III. The second, third, fourth, and fifth columns present the computation time of traditional RTCA-SCED, first component, DC-OPF, and second component, respectively. It can be observed from the second and third columns that the computation times of RTCA-SCED and the first component are of similar order. This is because the computational speeds of both of these approaches depend primarily on the number of critical contingencies identified by RTCA. This becomes especially clear after the third outage occurs (see last row, second and third columns of Table III). After this (third) outage, a relatively large number of violations were modeled as post-contingency branch overload constraints of SCED and iCA, which consequently increased the computation time of the traditional RTCA-SCED and the first component, respectively. It must also be noted that for this system, the computation time for SCED and iCA were obtained after the PTDFs lower than 0.02 were rounded off to 0. When this rounding was not done, due to the extremely high computational burden of the optimization problem for RTCA-SCED and the first component, the local memory of the solver became insufficient.

On a similar note, the computation times of DC-OPF and the second component are found to be comparable (see fourth and fifth columns of Table III). Both were less than 30 seconds for this system, which is at least an order of magnitude faster than the first component and RTCA-SCED. The high speed is primarily because the DC-OPF and rCA (used in the second component) do not model the computationally intensive post-contingency branch overload constraints. Furthermore, note that the optimization problems of the rCA and DC-OPF did not require any approximation of the PTDFs. However, the performance of the second component is superior in comparison to a simple DC-OPF because the former incorporates a relaxed criterion of security using the cut-set power transfer constraints (modeled inside rCA). *Thus, rCA is able to provide security against post-contingency cut-set saturation without significantly increasing the computational burden of the resulting optimization problem.*

#### 2) Real-time implementation of proposed methodology

Table IV presents the real-time application of the two-component methodology for the three outages described in the previous sub-section. The first column lists the sequence of events. Let the redispatch be implemented every 10 minutes. Then, it can be observed from Table III that the first component yields a result within 10 minutes for the first two outages, but the computation time is longer than 10 minutes after the third outage. Therefore, the redispatch solution from the first component should be implemented after the first and second outages occur, whereas the results from the second component should be used for redispatch after the third outage has occurred (see the second column, last row of Table IV). The third column presents the solution (generation cost and load-shed) obtained when one of the two components of the proposed methodology is implemented after every outage to mitigate the identified post-contingency violations. A summary of the observations made from the dispatch solution in Table IV is provided below.

- *Outage 1:* The generation redispatch (obtained using the first component) alone cannot mitigate the identified post-contingency violations. Therefore, 106 MW of load is shed at this stage. Therefore, the remaining load in the system becomes 67,003 (= 67,109-106) MW. The total generation fleet satisfies the power demand of 67,003 MW at the generation cost of $ 922.4k.
- *Outage 2:* Following the second event, the first component is implemented once more. To mitigate additional post-contingency violations, the generation cost for redispatch increases to $ 924.7k. The redispatch solution involves no additional load-shed, and so the load of 67,003 MW is satisfied by the new generation dispatch.
- *Outage 3:* Following the third event, the second component is implemented. The redispatch solution involves no additional load-shed indicating that the total generation now satisfies the power demand of 67,003 MW at a new generation cost of $ 923.2k. Note that the slight decrease in the generation cost from $ 924.7k to $ 923.2k is due to the relaxed security constraints of rCA (in the second component) compared to the more comprehensive security constraints of iCA (in the first component).

Finally, the last column presents the number of cascade-triggering contingencies contained in the solution. It is observed that for the listed sequence of events, the solution obtained from the proposed methodology does not contain any cascade triggering contingencies. *Therefore, this case-study illustrates the real-time implementation of the two-component methodology during multiple outages in large power systems.*

## V. CONCLUSION

This paper presents a two-component methodology that enhances the scope and speed, respectively, of static power system security assessment during multiple outage scenarios. The most salient aspect of the proposed methodology is its ability to secure the power system against saturated cut-sets occurring due to a potential contingency using a combination of graph theory and constrained optimization.

The first component of the proposed methodology combines the results from the FT algorithm and RTCA to create an integrated corrective action (iCA). The iCA initiates a comprehensive response to the violations detected by FT and RTCA to protect the system against saturated cut-sets as well as critical branch overloads.

The second component of the proposed methodology presents an alternative method that complements real-time power system operations during extreme event scenarios, when detailed network analysis tools such as the first component or traditional RTCA-SCED take longer time to generate a solution. Under such circumstances, by only employing the FT algorithm, a relaxed corrective action (rCA) is implemented that quickly secures the system against post-contingency cut-set saturation.

Multiple case-studies performed on the IEEE 118-bus system as well as a 2000-bus synthetic Texas system demonstrate the consistently good performance as well as scalability of the proposed approach. The results confirm that the proposed two-component methodology can successfully enhance power system security when multiple outages manifest successively in a region.

## ACKNOWLEDGEMENTS

The authors would like to acknowledge Dr. John Undrill from Arizona State University, for the valuable discussions and help during the course of this research.

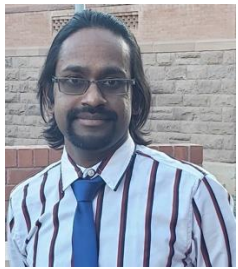

**Reetam Sen Biswas** (S'15) received the B.Tech. degree in electrical engineering from the West Bengal University of Technology, Kolkata, India, in 2016. He received the M.S. and Ph.D. degree in electrical engineering from Arizona State University, Tempe, AZ, USA in 2019 and 2021 respectively. His research interests include power system state estimation, contingency analysis, cascading failure analysis, power system security assessment and the integration of renewables into power systems.

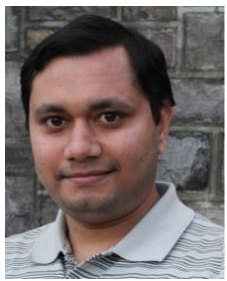

**Anamitra Pal** (S'11-M'15-SM'19) received his Bachelor of Engineering (B.E.) degree (summa) in electrical and electronics engineering from Birla Institute of Technology, Mesra, Ranchi (India) in 2008 and his M.S. and Ph.D. degrees in electrical engineering from Virginia Tech, Blacksburg in 2012 and 2014, respectively. He is now an Assistant Professor in the School of Electrical, Computer, and Energy Engineering at Arizona State University. Previously, from 2014-2016, he was an Applied Electrical and Computer Scientist in the Network Dynamics and Simulation Science Laboratory at the Biocomplexity Institute of Virginia Tech. He is the recipient of the 2018 Young CRITIS Award for his contributions to critical infrastructure resiliency, as well as the 2019 Outstanding IEEE Young Professional Award from the IEEE Phoenix Section. His current research interests include power system modeling, transient and dynamic stability analysis, critical infrastructure resiliency assessment, and wide area measurements-based protection, monitoring, and control.

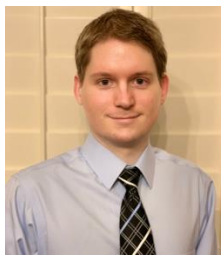

**Trevor Werho** (S'14-M'15) received the degrees of B.S.E., M.S., and Ph.D. degrees in electrical engineering from Arizona State University, Tempe, AZ, USA, in 2011, 2013, and 2015, respectively. He completed working as a post-doctoral scholar at Arizona State University researching wind and solar forecasting in 2021. His research interests include power system analysis and the integration of renewables into power systems.

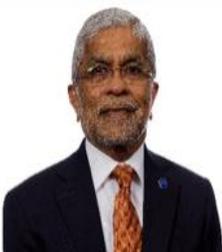

**Vijay Vittal** (S'78-M'82-SM'87–F'97) received the B.E. degree in electrical engineering from the B.M.S. College of Engineering, Bangalore, India, in 1977, the M. Tech. degree from the Indian Institute of Technology, Kanpur, India, in 1979, and the Ph.D. degree from Iowa State University, Ames, in 1982.

He is a Regents' Professor and the Ira A. Fulton Chair Professor in the Department of Electrical, Computer, and Energy Engineering at Arizona State University, Tempe, AZ. Dr. Vittal is a member of the National Academy of Engineering. He is a recipient of the IEEE PES Outstanding Power Engineering Educator award in 2000, the IEEE Herman Halperin T&D Field Award in 2013, and the IEEE PES Prabha S. Kundur Power System Dynamics and Control Award in 2018.